\documentclass{PoS}

\usepackage{url}
\usepackage{amsmath}
\usepackage{color}
\usepackage[absolute]{textpos}

\title{nPDF constraints from the Large Hadron Electron Collider}

\ShortTitle{nPDF constraints from the LHeC}

\author{\speaker{Ilkka Helenius}$^{a}$, Hannu Paukkunen$^{bcd}$ and Nestor Armesto$^{d}$\\
\llap{$^a$}Department of Astronomy and Theoretical Physics, Lund University, S\"{o}lvegatan 14A, SE-223 62 Lund, Sweden\\
\llap{$^b$}University of Jyvaskyla, Department of Physics, P.O. Box 35, FI-40014 University of Jyvaskyla, Finland\\
\llap{$^c$}Helsinki Institute of Physics, P.O. Box 64, FI-00014 University of Helsinki, Finland\\
\llap{$^d$}IGFAE, Universidade de Santiago de Compostela, E-15782 Galicia, Spain\\
E-mail: \email{ilkka.helenius@thep.lu.se}, \email{hannu.t.paukkunen@jyu.fi}, \email{nestor.armesto@usc.es}}

\abstract{
An updated analysis regarding the expected nuclear PDF constraints from the future Large Hadron Electron Collider (LHeC) experiment is presented. The new study is based on a more flexible small-$x$ parametrization which provides less biased uncertainty estimates in the region where there are currently no data constraints. The effect of the LHeC is quantified by directly including a sample of pseudodata according to the expected precision of this planned experiment. As a result, a significant reduction of the small-$x$ uncertainties in sea quarks and gluons is observed.
}

\FullConference{XXIV International Workshop on Deep-Inelastic Scattering and Related Subjects\\
         11-15 April, 2016\\
         DESY Hamburg, Germany}

\begin{document}

\begin{textblock}{2}(13.0,0.8)
\mbox{}\hfill LU TP 16-34\\
\mbox{}\hfill June 2016
\end{textblock}

\section{Introduction}

The Large Hadron Electron Collider (LHeC) project consists on the construction of a new electron accelerator providing an electron beam that would collide with the hadron beam from the Large Hadron Collider (LHC). The design also allows a synchronous operation of e-p collisions at the LHeC and p-p collisions at the LHC, as well as replacing the proton beam with heavy ions. A detailed description about the design can be found in the conceptual design report \cite{AbelleiraFernandez:2012cc}.

Accurately determined parton distribution functions (PDFs) are essential to all perturbative QCD (pQCD) based theory predictions. This holds also for nuclear collisions, where the nuclear modifications of the PDFs are needed to obtain a precise pQCD baseline for heavy-ion physics at the LHC, and also for future high-energy colliders, such as the Future Circular Collider (FCC) \cite{Dainese:2016gch}. In this talk we focus on the current status of the nuclear PDFs (nPDFs), and quantify the potential of  e-Pb collisions at the LHeC to improve the accuracy of the future nPDF analyses. Other interesting e-$A$ physics are described e.g. in Refs.~\cite{AbelleiraFernandez:2012cc, Helenius:2015xda}.

\section{New baseline nPDF fit}

The nPDFs can be determined via a similar global analysis as the free proton PDFs. The scale evolution is given by the DGLAP equations, but a non-perturbative input function $f(x,Q_0^2)$ at the initial scale $Q_0^2$ of the fit is required for the $x$ dependence. The ansatz for $f(x,Q_0^2)$ should include enough freedom to capture all the relevant features of data used in the analysis but, on the other hand, keep the number of parameters reasonable to ensure the convergence of the fit and not to be sensitive to the fluctuations of the individual data points.

The data used in the current fits comes from deep inelastic scattering (DIS), Drell-Yan dilepton production and inclusive pion production in d-Au collisions at RHIC. Since most of the data are from fixed-target experiments, the limited collision energy restricts the kinematic reach of the data, as discussed e.g. in Ref.~\cite{Helenius:2015xda}. Therefore, the current fits \textsc{ncteq15}~\cite{Kovarik:2015cma}, \textsc{ka15}~\cite{Khanpour:2016pph}, \textsc{dssz}~\cite{deFlorian:2011fp}, \textsc{eps09}~\cite{Eskola:2009uj} and \textsc{hkn07}~\cite{Hirai:2007sx} include data that are restricted to $x~\gtrsim 5\cdot 10^{-3}$ and $Q^2\lesssim 100~\mathrm{GeV^2}$. The recent dijet, charged hadron and $W^{\pm}/Z$ data from p-Pb collisions at the LHC provide some additional constraints for the nPDF analyses \cite{Armesto:2015lrg}, but in practice do not increase the small-$x$ reach. With the LHeC, the kinematic reach of e-$A$ DIS cross section could be extended by more than three orders of magnitude in both $x$ and $Q^2$ providing a kinematic reach comparable to the current proton PDF fits.

\begin{figure}[thb!]
\centering
\includegraphics[clip, trim=0pt 12pt 0pt 34pt, width=0.49\textwidth]{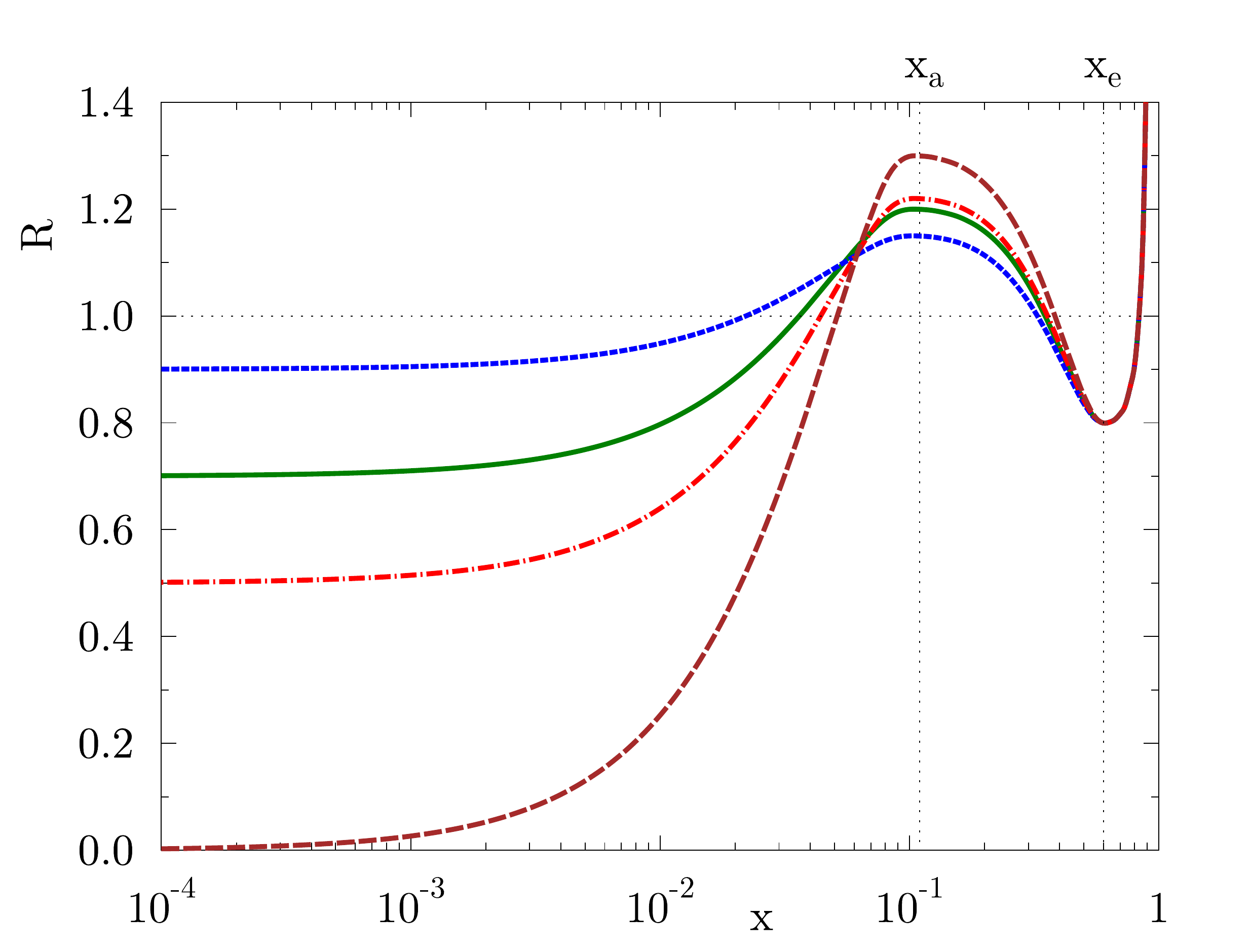}
\includegraphics[clip, trim=0pt 12pt 0pt 34pt, width=0.49\textwidth]{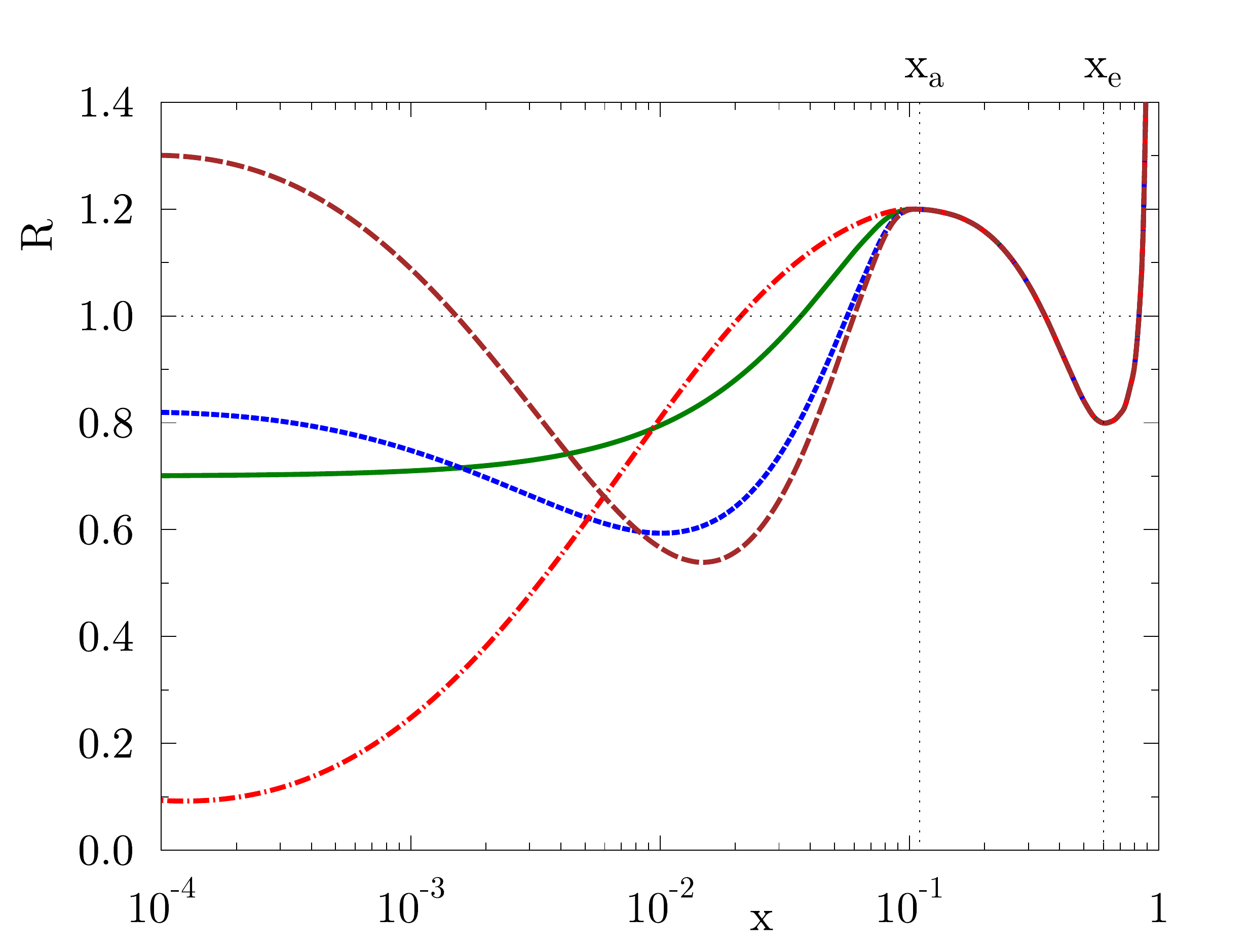}
\caption{The fit function behaviour from the \textsc{eps09}-style ansatz (left) and from the new ansatz with more freedom at small values of $x$ used in this study (right).}
\label{fig:epsForm}
\end{figure}

The nPDF uncertainties are usually quantified via the Hessian method \cite{Pumplin:2001ct}, which allows a simple way to study how the uncertainties propagate into physical observables. Within this method the parameters (in a diagonalized basis) are allowed to vary within some $\Delta \chi^2$ value (here, $\Delta \chi^2=17$). The uncertainty estimates are always tied to the functional form of $f(x,Q_0^2)$, especially in the region where there are no guidance from the data. In the \textsc{eps09} analysis \cite{Eskola:2009uj} the ratios $R$ of parton densities in a bound proton inside a nucleus over that in a free proton were parametrized with a piecewise function,
\begin{equation}
R^{\rm EPS09}(x) = 
\left\{
\begin{array}{lc}
a_0 + \left(a_1+a_2x \right) \left( e^{-x}-e^{-x_a} \right), & x \leq x_a \\
b_0 + b_1x + b_2x^2 + b_3x^3, & x_a \leq x \leq x_e \\
c_0 + \left(c_1-c_2x \right) \left(1-x\right)^{-\beta}, & x_e \leq x \leq 1
\end{array}
\right. \, . 
\end{equation}
This functional form does not allow much freedom for the small-$x$ behaviour, as shown in the left-hand panel of Figure \ref{fig:epsForm} --- the function at small $x$ is monotonic. Theoretically, this kind of behaviour appears reasonable but since global fits should rather indicate the uncertainty originating from the used data, a more flexible form should be used to study the realistic potential of the LHeC. The small-$x$ form used in this study is
\begin{equation}
R(x \leq x_a) = a_0 + a_1(x-x_a)^2 + \sqrt{x}(x_a-x) \left[a_2\log\left(\frac{x}{x_a}\right)+a_3\log^2\left(\frac{x}{x_a}\right)+a_4\log^3\left(\frac{x}{x_a}\right) \right],
\label{eq:Rnew}
\end{equation}
which provides much more freedom for the small-$x$ behaviour reducing the bias due to the ansatz significantly. This is illustrated in the  right-hand panel of Figure \ref{fig:epsForm}, showing the $R(x<x_a,Q_0^2)$ with a few different possible parameter values.

\begin{figure}[thb!]
\centering
\includegraphics[clip, trim=0pt 7pt 0pt 20pt, width=0.85\textwidth]{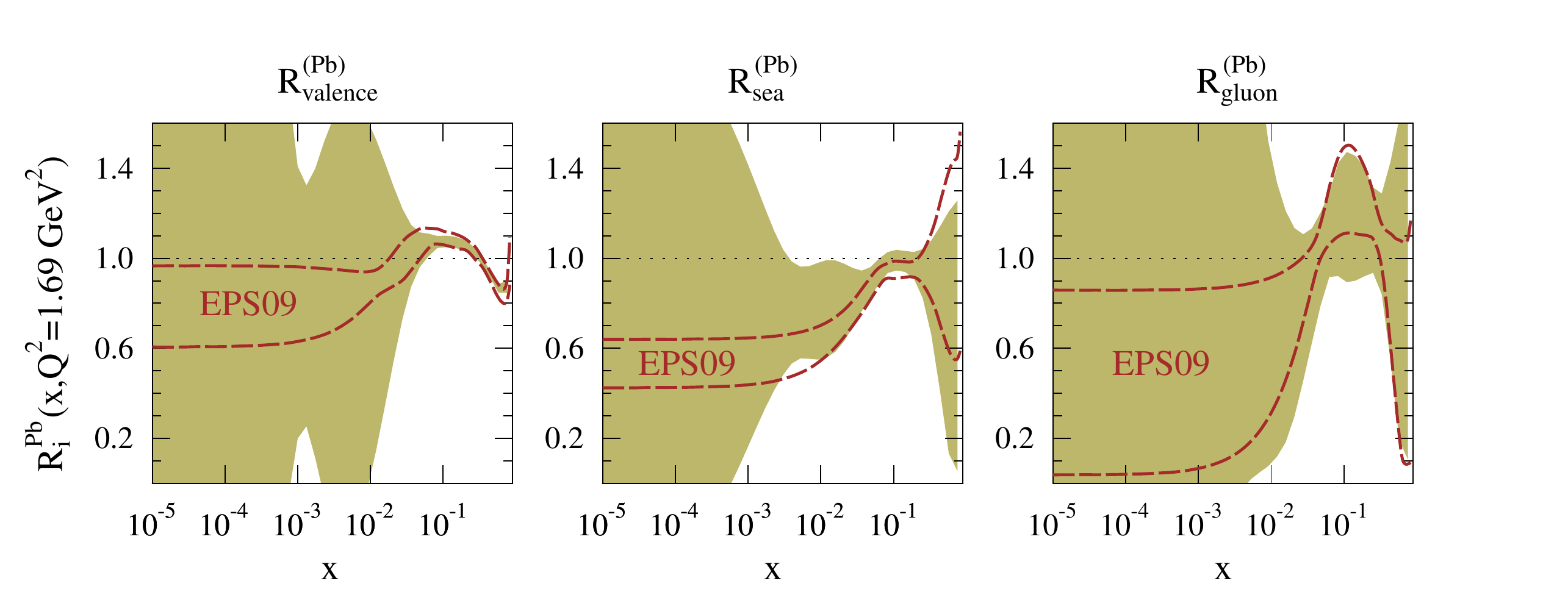}\\
%\vspace{-12pt}
\includegraphics[clip, trim=0pt 10pt 0pt 55pt, width=0.85\textwidth]{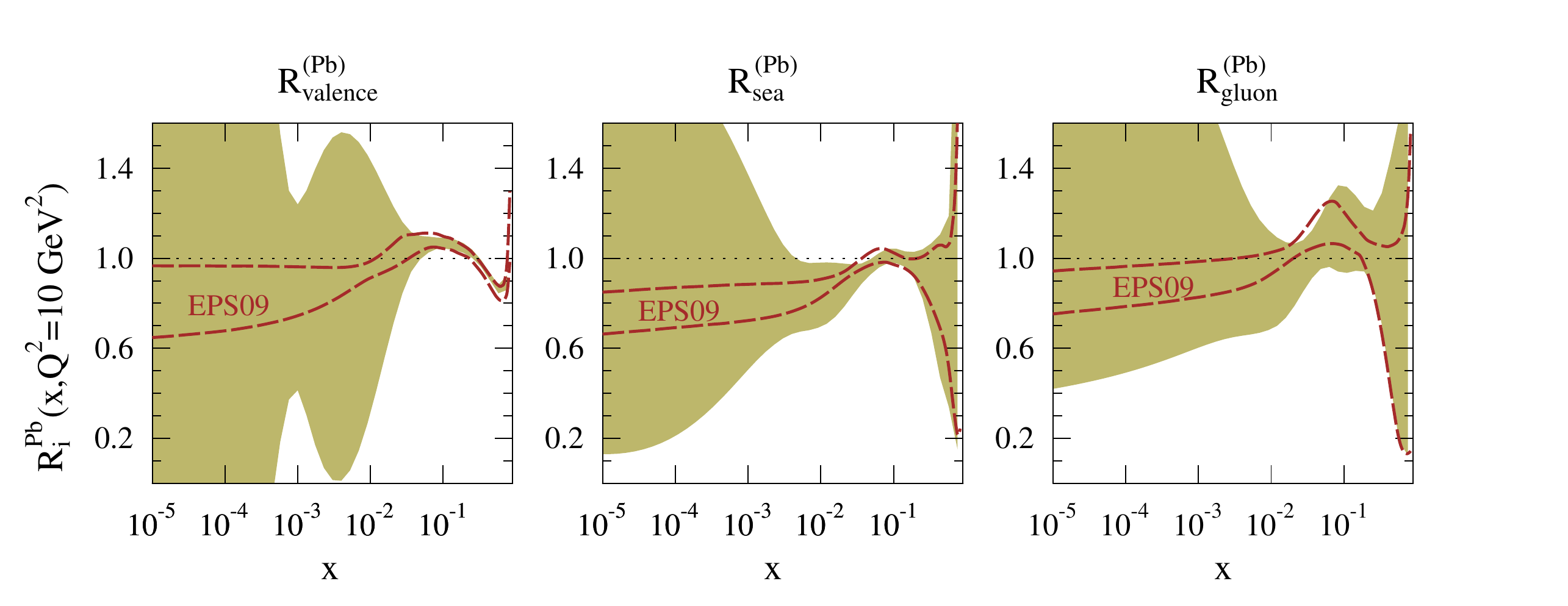}
\caption{The nPDF modifications for valence (left) and sea (middle) quarks, and gluons (right) at $Q^2 = 1.69~\mathrm{GeV^2}$ (upper panels) and $Q^2 = 10~\mathrm{GeV^2}$ (lower panels). The dashed lines show the uncertainty according to the \textsc{eps09} analysis and the colour band the result of the new baseline fit with the more flexible low-$x$ parametrization.}
\label{fig:RepsNew}
\end{figure}

To have a more realistic estimate for the current uncertainties, a new baseline fit was performed. The setup for the baseline fit was very similar as in the \textsc{eps09} analysis, but now including the more flexible form from Eq.~(\ref{eq:Rnew}) at low $x$. The new baseline fit reflects the lack of control at small values of $x$ from the data by providing vastly larger uncertainties than the original \textsc{eps09} analysis, as shown in Figure \ref{fig:RepsNew}, even though the data constraints are the same. 

\section{Fit to pseudodata}

\begin{figure}[thb!]
\centering
\includegraphics[clip, trim=0pt 40pt 0pt 50pt, width=\textwidth]{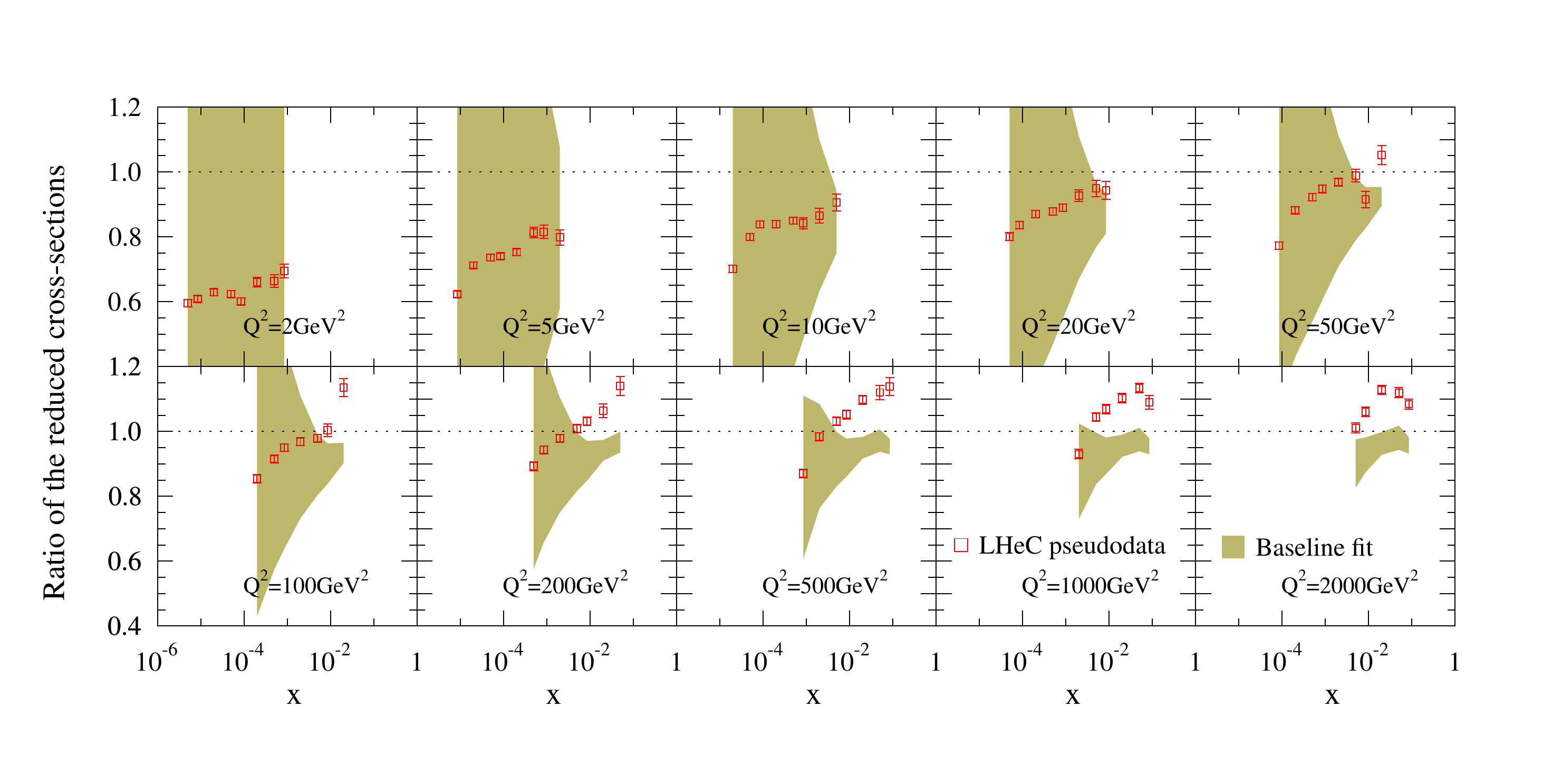}
\includegraphics[clip, trim=0pt 44pt 0pt 50pt, width=\textwidth]{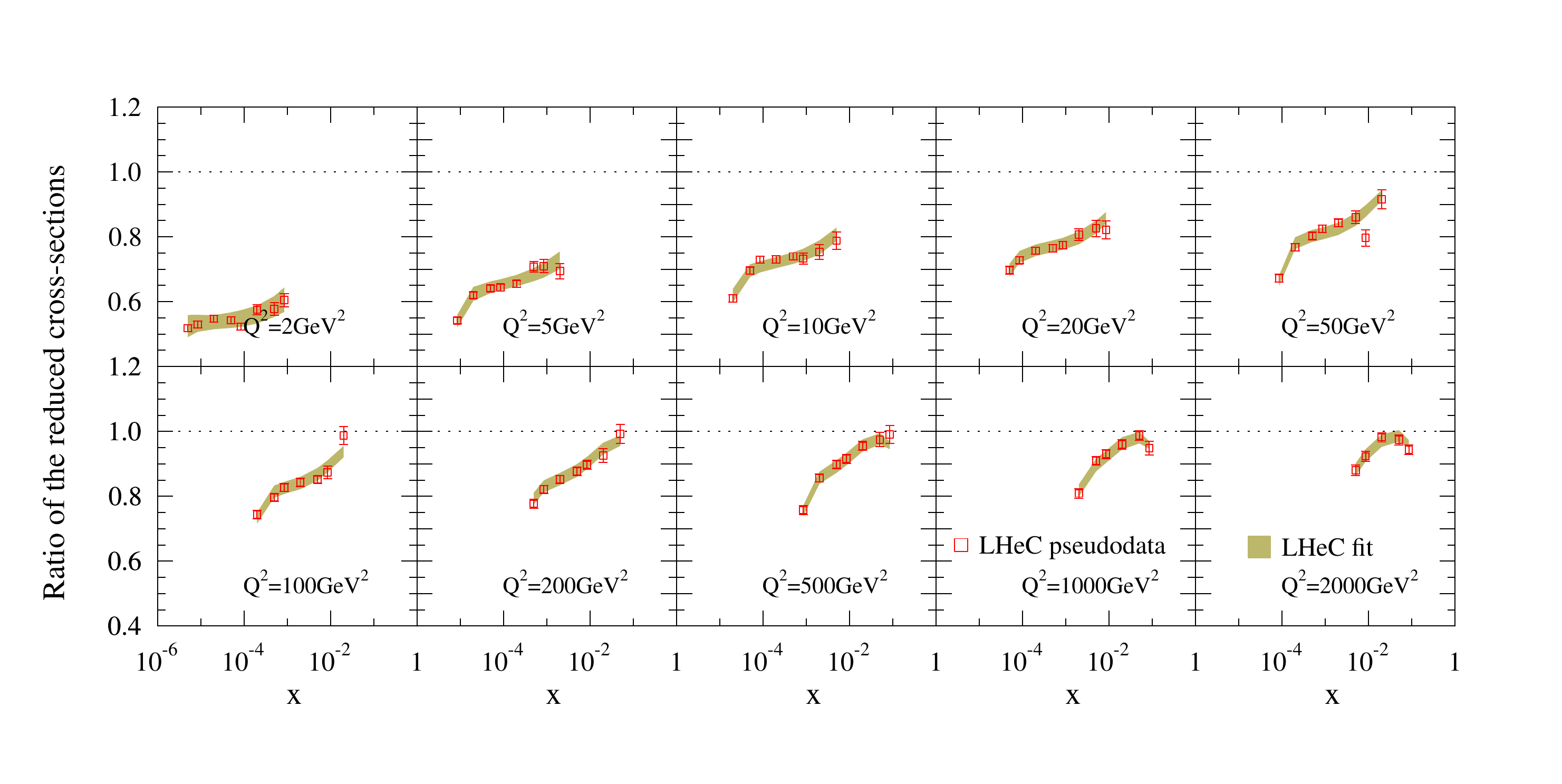}
\caption{Ratios of the reduced NC DIS cross section between e-Pb and e-p collisions. The generated pseudodata are shown with red markers and compared to the baseline fit (upper panels) and to a new fit using the LHeC pseudodata (lower panels). The fit uncertainties are shown with yellowish colour bands.}
\label{fig:sigmaNC}
\end{figure}

The impact of the LHeC pseudodata on the nPDF uncertainties was studied by performing a new nPDF analysis using pseudodata generated according to LHeC expectations, but otherwise keeping the details as in the baseline fit. The pseudodata were generated for neutral current (NC) DIS assuming per-nucleon luminosities $\mathcal{L}_{\mathrm{ep}} = 10~\mathrm{fb}$ and $\mathcal{L}_{\mathrm{ePb}} = 1~\mathrm{fb}$. The kinematical window studied was $10^{-5}<x<1$ and $2<Q^2<10^5~\mathrm{GeV^2}$ and the data were based on the \textsc{eps09} nPDFs, including also realistic fluctuations with the cited luminosities \cite{MaxKlein}. A comparison between the baseline fit and the generated pseudodata is shown in Figure~\ref{fig:sigmaNC} displaying ratios of reduced cross sections between e-Pb and e-p collisions. Figure~\ref{fig:sigmaNC} shows also the corresponding comparison after including the pseudodata into the fit. A drastic reduction of the nPDF uncertainties is observed once the pseudodata are included to the analysis.

\begin{figure}[thb!]
\centering
\includegraphics[clip, trim=0pt 10pt 0pt 20pt, width=0.85\textwidth]{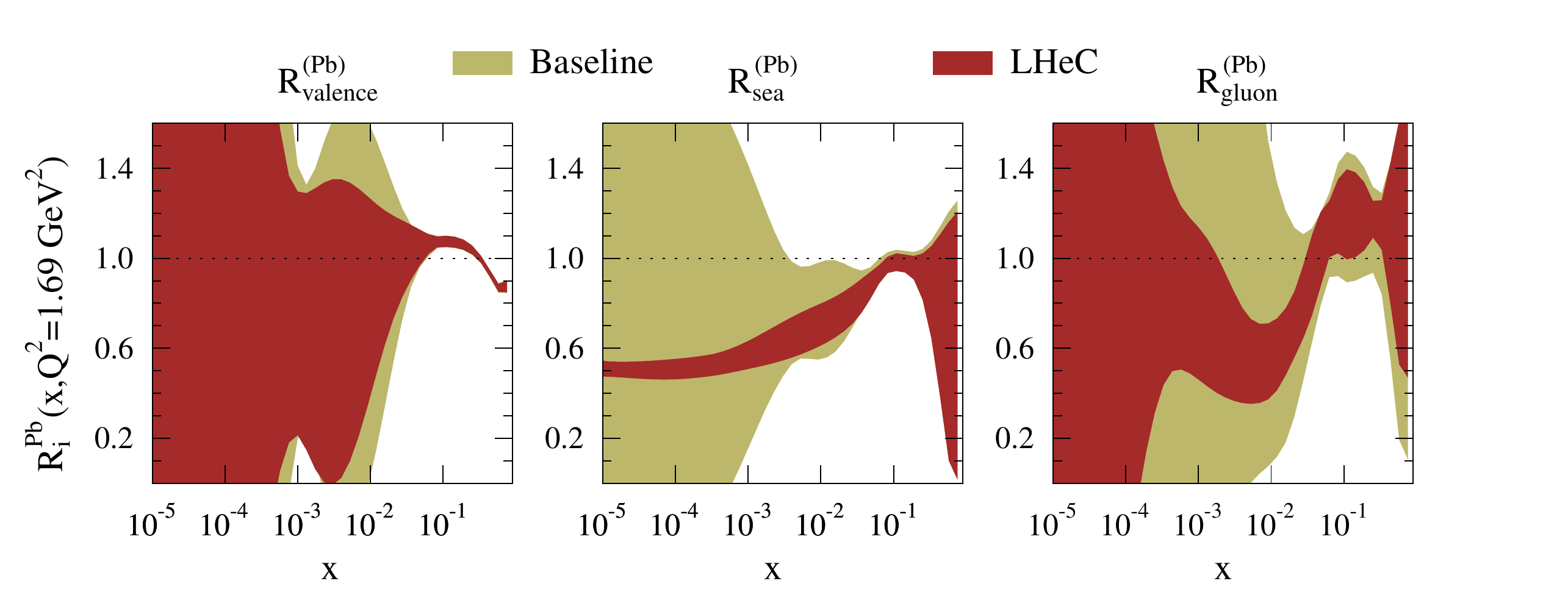}
\includegraphics[clip, trim=0pt 10pt 0pt 20pt, width=0.85\textwidth]{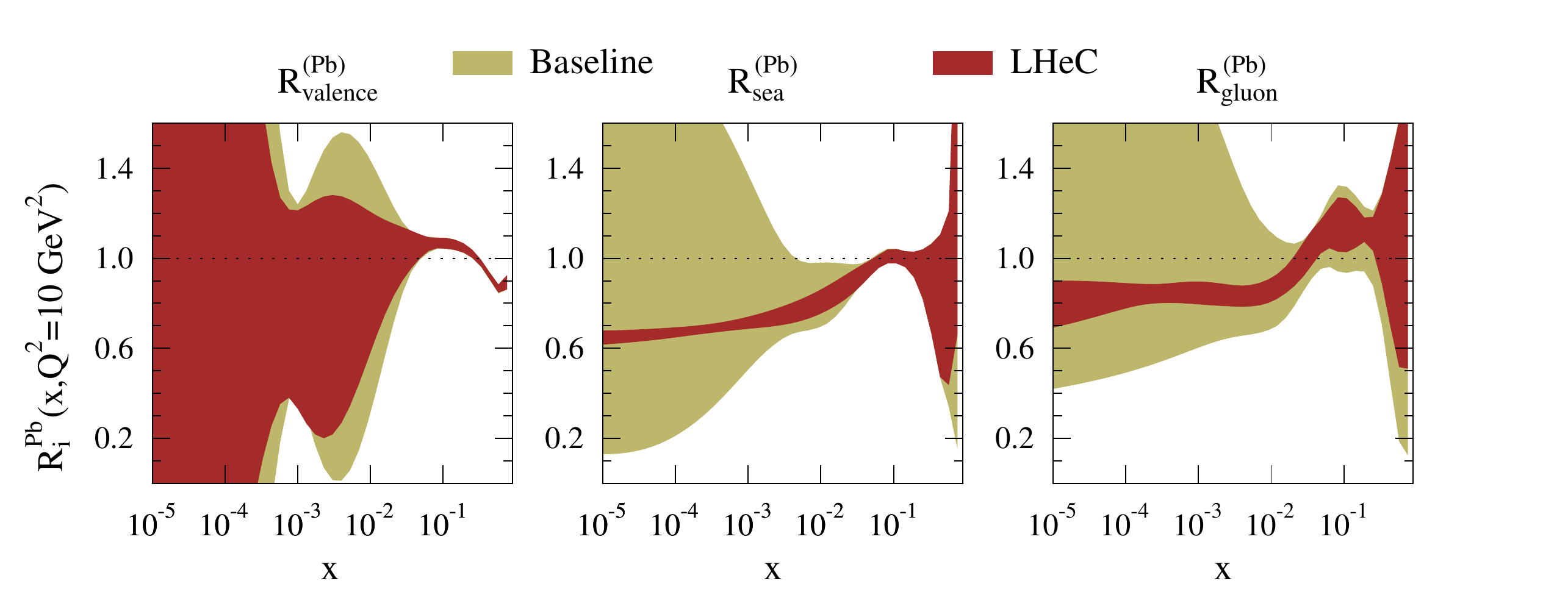}
\caption{The nuclear modification of the PDFs for valence quarks (left), sea quarks (middle) and gluons (right) for two scales $Q^2=1.69~\mathrm{GeV^2}$ (upper panels) and $Q^2=10~\mathrm{GeV^2}$ (lower panels). The baseline fit is shown as yellowish colour bands and the fit including also LHeC pseudodata as dark red bands.}
\label{fig:newR}
\end{figure}

To study the improvements in more detail, Figure~\ref{fig:newR} compares the nPDF uncertainties for valence quarks, sea quarks and gluons from the baseline fit and from the new fit including also the pseudodata. The large-$x$ valence quarks are already well constrained by the current data but the improvement is significant for the small-$x$ sea quarks and gluons. Improvement like this would substantially solidify the pQCD baseline for the heavy-ion physics at the LHC, and also at the FCC, already from $p_T\sim 3~\mathrm{GeV/c}$ on.

\section{Summary}

The impact of the NC DIS data from the LHeC to the nPDFs has been studied. To obtain a more realistic estimates than the present nPDF uncertainties, a new baseline fit was performed applying a more flexible form for the small-$x$ behaviour that is included in the current fits. Due to the new parametrization, the baseline analysis possesses significantly larger small-$x$ uncertainties than the current public analyses. A set of pseudodata was generated according to the expected LHeC specifications, and a new fit including the pseudodata was performed and the uncertainties were compared to the baseline fit without the pseudodata. The results show a substantial reduction of small-$x$ uncertainties for sea quarks and gluons, improving the accuracy of pQCD baseline for heavy-ion collisions in the kinematic region relevant for the LHC and the FCC.

\acknowledgments
This research was supported by the European Research Council grant HotLHC ERC-2011-StG-279579 and by Xunta de Galicia (Conselleria de Educacion) --- H.~P. and N.~A. are part of the Strategic Unit AGRUP2015/11.  N. A. is supported by the People Programme (Marie Curie Actions) of the European Union Seventh Framework
Programme FP7/2007-2013/ under REA grant agreement \#318921 and Ministerio
de Ciencia e Innovaci\'onn of Spain under project FPA2014-58293-C2-1-P. I~.H. has been supported by the MCnetITN FP7 Marie Curie Initial Training Network, contract PITN-GA-2012-315877 and has received funding from the European Research Council (ERC) under the European Union Horizon 2020 research and innovation programme (grant agreement No 668679).

\bibliographystyle{JHEP}
\bibliography{eALHeC}

\end{document}